\documentclass[letterpaper, 10 pt, conference]{ieeeconf}  

\IEEEoverridecommandlockouts                              
\usepackage{graphics} 
\usepackage{subcaption}

\usepackage{xcolor}
\usepackage{epsfig} 
\usepackage{amsmath, bm}
\usepackage{amssymb}  
\usepackage{arydshln}
\usepackage{amsfonts}

\newtheorem{definition}{Definition}
\newtheorem{remark}{Remark}

\usepackage[noadjust]{cite}
\title{\LARGE \bf
Exploring the use of deep learning in task-flexible ILC*
}

\author{Anantha Sai Hariharan Vinjarapu, Yorick Broens, Hans Butler, {and Roland T\'{o}th}
\thanks{*This work has received funding from the ECSEL Joint Undertaking under grant agreement No 875999 and from the European Union within the framework of the National Laboratory for Autonomous Systems (RRF-2.3.1-21.2022-00002).}
\thanks{A.S.H.Vinjarapu, Y.Broens, R.T\'oth {and H.Butler} are with the Department of Electrical Engineering, Eindhoven University of Technology, Eindhoven, The Netherlands. R.T\'oth is also affiliated with the Systems and Control Laboratory, Institute for Computer Science and Control, Hungary. {H.Butler is also affiliated with ASML, Veldhoven, The Netherlands}. Email: {\tt\small y.l.c.broens@tue.nl}.}
}

\begin{document}
\maketitle
\thispagestyle{empty}
\pagestyle{empty}
\begin{abstract}

Growing demands in today's industry results in increasingly stringent performance and throughput specifications. For accurate positioning of high-precision motion systems, feedforward control plays a crucial role. Nonetheless, conventional model-based feedforward approaches are no longer sufficient to satisfy the challenging performance requirements. An attractive method for systems with repetitive motion tasks is iterative learning control (ILC) due to its superior performance. However, for systems with non-repetitive motion tasks, ILC is {generally} not applicable, {despite of some recent promising advances}. In this paper, we aim to explore the use of deep learning to address the task flexibility constraint of ILC. For this purpose, a novel Task Analogy based Imitation Learning (TAIL)-ILC approach is developed. To benchmark the performance of the proposed approach, a
simulation study is presented which compares the TAIL-ILC to classical model-based feedforward strategies and existing learning-based approaches, such as neural network based feedforward learning.



\end{abstract}
\section{Introduction}
\label{intro}
High-precision positioning systems are essential components in modern manufacturing machines and scientific equipment, see \cite{Butler,411117,HEERTJES20161, 6225187}. To ensure high-throughput and high-accuracy position tracking, a two-degree-of-freedom controller structure, consisting of a feedback controller and a feedforward controller, is commonly utilized, see \cite{1223174,Oomen,190Steinbuch}.  The feedback controller maintains closed-loop stability and disturbance rejection, while the feedforward controller is primarily responsible for achieving optimal position tracking performance, see \cite{articleHH}.
Nonetheless, with the increasingly stringent demands in contemporary industry, conventional model-based feedforward techniques, e.g. \cite{oomen2020model}, are no longer adequate to meet the desired performance specifications, thus necessitating for alternative feedforward approaches.

\emph{Iterative Learning Control} (ILC), see \cite{ahn2007iterative}, has emerged as a viable choice for feedforward control in motion systems that execute recurring tasks, enabling accurate position tracking. Despite its advantages, ILC exhibits significant limitations. Primarily, ILC is dependent on the assumption that the tracking error recurs from one iteration to the next, limiting its general applicability. Additionally, conventional ILC performance is constrained to a single task, see \cite{BLANKEN2016213}. 



Several studies have attempted to address the task flexibility limitations of ILC by drawing on concepts from machine learning and system identification, as reported in the literature \cite{HulstMSc,5559384,Bosma-msc}. However, the findings from the related literature suggest that there exists a trade-off between the achievable position tracking performance and the degree of deviation from the core principle of ILC, i.e., direct iterative manipulation of signals. Instead of compromising local ILC performance to enhance task flexibility, the aim is to develop a learning-based feedforward strategy that can deliver superior position tracking performance regardless of the severity of the variation of the compensatory signal across tasks.
Such an ILC variant can be imagined to make use of imitation learning in order to mimic the behaviour of conventional ILC policies generalized over multiple trajectories. 

This paper introduces a novel approach to ILC, termed Task Analogy based Imitation Learning (TAIL)-ILC, from a data science perspective. By acquiring spatial feature analogies of the trajectories and their corresponding control signals, performance of conventional ILC policies  can be replicated. To facilitate efficient network training, abstract lower-dimensional representations of signals are utilized. This approach offers numerous benefits in terms of training and prediction time efficiency, utilization of large datasets, and high sampling rate handling. The resulting feedforward controller comprises an encoding policy, a learning policy, and a decoding policy arranged in a cascade interconnection. Dual principal component analysis (DPCA), a standard linear dimensionality reduction technique, is utilized for the integration of the encoding and decoding policies, while a deep neural network is employed for the learning policy.

 
 The main contributions of this paper are:
 \vspace*{-1mm}
 \begin{itemize}
     \item [(C1)] {A novel TAIL-ILC approach that tackles the task extension problem of ILC via learning spatial feature analogies of trajectories and their {compensation} 
     signals, enabling direct imitation of 
     ILC policies.} 
     \item [(C2)] 
     {An efficient implementation strategy is devised for the learning-based feedforward controller in terms of constructing it through the cascade interconnection of an encoder, a deep neural network, and a decoder.}
 \end{itemize}
  \vspace*{-1mm}
 This paper is organized as follows. First, the problem formulation is presented in Section \ref{Section_Problem_formulation}. Next, Section \ref{Section_Conventional_Feedforward} 
 presents the proposed novel TAIL-ILC approach which aims  at generalizing ILC performance across various tasks through imitation learning strategies.  Section \ref{Section_SimulationStudy} provides a simulation study of the proposed approach with respect to existing feedforward strategies using a high-fidelity model of a moving-magnet planar actuator. In Section \ref{TAILPERSPE}, detailed comparison between the proposed TAIL-ILC approach and neural-network-based feedforward strategies is presented. Finally, conclusions on the proposed approach are presented in Section \ref{Section_Conclusions}.
 
 \section{Problem statement}
 \label{Section_Problem_formulation}
 \subsection{Background}
 \label{PF}

\begin{figure}[t]
    \centering
    \includegraphics[trim={0cm 0cm 0cm 0cm},width=.7\linewidth]{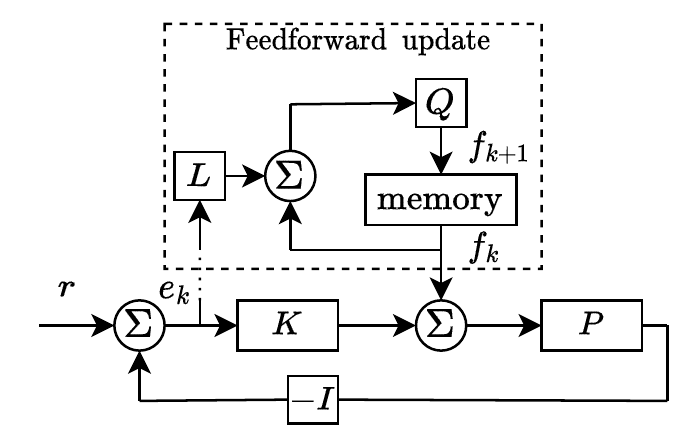} \vspace{-3mm}
     \caption{Control structure with the conventional ILC configuration.}
     \label{fig:ILC_interconnection} \vspace{-10mm}
 \end{figure}

 Consider the conventional frequency domain ILC configuration illustrated by Figure \ref{fig:ILC_interconnection}, where $P\in \mathcal{R}^{n_\mathrm{y} \times n_\mathrm{u}}$ corresponds to the proper {transfer matrix representation} of a \emph{discrete time} (DT) \emph{linear-time-invariant} (LTI) 
\emph{multiple-input multiple-output} (MIMO) plant {with $\mathcal{R}$ denoting the set of real rational functions in the complex variable $z\in\mathbb{C}$.} Furthermore, the proper $K\in \mathcal{R}^{n_\mathrm{u} \times n_\mathrm{y}}$ represents a LTI stabilizing DT feedback controller, which is typically constructed using rigid-body decoupling strategies, see \cite{Steinbuch2013}.
The aim of the conventional frequency domain ILC framework is to construct an optimal feedforward policy $f$, which minimizes the position tracking error $e$ in the presence of the motion trajectory $r$. Under the assumption that the reference trajectory is trial invariant, the error propagation per trial $k \in \mathbb{N}_{\geq 0}$ is given by:
\vspace*{-.2cm}
 \begin{equation}
     e_k = Sr-Jf_k,
          \label{ILC_Formula_1}
 \end{equation}
 
 \vspace*{-.1cm}

 \noindent where $S=(I+PK)^{-1}$ and $J=(I+PK)^{-1}P$. Generally, the update law for the feedforward policy is in accordance with the procedure outlined in \cite{blanken2019multivariable}:
 \vspace*{-.2cm}
 \begin{equation}
     f_{k+1} = Q\left(Le_k+f_k \right),
     \label{ILC_Formula_2}
 \end{equation}
 
 \vspace*{-.1cm}
 
 \noindent where $L\in \mathcal{RL}_{\infty}^{n_\mathrm{u} \times n_\mathrm{y}} $ is a learning filter and $Q \in \mathcal{RL}_\infty^{n_\mathrm{u} \times n_\mathrm{u}}$ denotes a robustness filter {with $\mathcal{RL}_\infty$ corresponding to set of real rational functions in $z$ that have bounded singular value on the unit circle $\mathbb{D}=\{ e^{\mathrm{i} \omega} \mid \omega \in [0,2\pi] \}$, i.e., finite $\mathcal{L}_\infty (\mathbb{D} )$ norm.} Both $L$ and $Q$ are required to be designed for the ILC task at hand. Furthermore, by combining (\ref{ILC_Formula_1}) and (\ref{ILC_Formula_2}), the progression of the error and feedforward update is reformulated as: 
 \vspace*{-.2cm}
 \begin{subequations}  \label{ILC_Formula_3}
 \begin{align}
     e_{k+1} &= (I-JQJ^{-1} )Sr + JQ( J^{-1} -L ) e_k,\\ 
     f_{k+1} &= QLSr + Q(I-LJ)f_k,
     \end{align} 
     \end{subequations}
     
     \vspace*{-.1cm}

 \noindent which can be reduced to: 
 \vspace*{-.2cm}
 \begin{subequations}  \label{ILC_Formula_4}
 \begin{align}
     e_{k+1} &= (I-Q)Sr+Q(I-JL)e_k, \\
      f_{k+1} &= QLSr + Q(I-LJ)f_k,
  \end{align} 
     \end{subequations}
 
 \vspace*{-.1cm}
 \noindent under the assumption that $Q$ is diagonal and $J$ is approximately diagonal, which holds in case of rigid-body decoupled systems. 
 
 From (\ref{ILC_Formula_4}), several observations can be made. First, it can be observed that the contribution of $r$ to the position tracking error is dependent on the robustness filter $Q$, which is optimally chosen as identity to negate the contribution of the reference trajectory towards the tracking error. Secondly, learning filter $L$ aims to minimize the criterion $\|Q(I-JL) \|_\infty <1$, {where $\|\centerdot\|_\infty$ stands for the $\mathcal{H}_\infty$ norm,} such that the tracking error is steered to zero, which is optimally achieved when $L=J^{-1}$. Note that these assumptions on $Q$ and $L$ yield the optimal feedforward update $f_{k+1} = P^{-1}r$, which results in perfect position tracking. Moreover, when the convergence criterion is satisfied, the limit policies, i.e. $e_\infty = \lim_{k \rightarrow \infty }e_k$, $f_\infty = \lim_{k \rightarrow \infty }f_k$, correspond to:  \vspace*{-.2cm}
 \begin{subequations}  \label{ILC_Formula_5}
 \begin{align}
         e_\infty &= \bigl(I-J\bigl(I-Q(I-LJ)\bigr)^{-1}QL\bigr)Sr, \\
         f_\infty &= \left(I-Q(I-LJ)\right)^{-1}QLSr,
\end{align}
     \end{subequations}
 \vspace*{-.1cm}

In spite of its simplicity and efficacy, the conventional ILC is hindered by significant limitations, the most notable of which is its confinement to a single task. Consequently, its practical utility is restricted to particular types of machinery. 

 \subsection{Problem formulation}
 The aim of this paper is to address the challenge of augmenting the task-flexibility of the conventional ILC by utilizing an imitation learning based controller. This approach facilitates the generalization of the optimal feedforward policy, created by the conventional ILC, for a wider range of motion profiles. The primary objective of this paper is to devise a feedforward controller that employs a learning-based mechanism, which satisfies the following requirements:
 \begin{itemize}
     \item [(R1)] The learning-based feedforward approach enables the generalization of the performance of the conventional ILC across multiple trajectories.
     \item [(R2)] 
     The scalability of the learning-based feedforward approach is imperative for its implementation in systems with a high sampling rate.
 \end{itemize}

\section{TAIL-ILC}
\label{Section_Conventional_Feedforward}
\vspace*{-.2cm}
\subsection{Approach}
\label{3A_approach}
For a given dynamic system with a proper discrete transfer function $G\in\mathcal{R}$ under a sampling time $T_\mathrm{s} \in \mathbb{R}_{+}$, 
a reference trajectory $r$ of duration $T=n_\mathrm{d} T_\mathrm{s}$ seconds can be defined as 
\vspace*{-.3cm}
\begin{equation}
r = \begin{bmatrix}r(0) & \cdots & r(n_\mathrm{d}) \end{bmatrix}^\top ,
\end{equation}

\vspace*{-.1cm}
\noindent where $n_\mathrm{d}$ corresponds to the length of the signal in DT. This reference trajectory for example can correspond to a $n^{th}$ order motion profile. A trajectory class $C \subset  \mathbb{R}^{n_\mathrm{d} \times n_\mathrm{t}}$ is defined as a collection of reference trajectories such that each trajectory shares certain prominent spatial features (motion profile order, constant velocity interval length, etc.) with the others, where $n_\mathrm{t}$ is the number of trajectories:
\vspace*{-.2cm}
\begin{equation}
C = \{r_1,r_2,r_3 . . . . . ,r_{n_\mathrm{t}}\}.
\end{equation}

\vspace*{-.2cm}

\noindent 
%

 Given a specific combination of the $L$ and $Q$ filters, consider that an ILC policy $\pi^*$ exists which maps a given reference trajectory $r$ to the optimal feedforward compensation signal $f^*$, see  \eqref{ILC_Formula_5}. This can be formally expressed as:
 \vspace*{-.4cm}
\begin{equation} \pi^*: r_i \rightarrow f^*_i. \end{equation}

\vspace*{-.1cm}

\noindent 
Henceforth, $\pi^*$ shall be denoted as the expert policy, which is equipped with learning and robustness filters established through a process model. Our objective is to formulate an optimal student policy $\pi_\mathrm{s}^*$ that approximates the performance of the optimal policy $\pi^*$ over a set of trajectories from the pertinent trajectory class. To this end, we endeavor to determine $\pi_\mathrm{s}^*$ as a solution to the optimization problem:
\vspace*{-.1cm}
\begin{equation}
\label{mes}
\pi_\mathrm{s}^* = \underset{\pi_\mathrm{s}}{\arg\min}\ \eta(\pi^*(r_i),\pi_\mathrm{s}(r_i)),\quad \forall i \in [1,n_\mathrm{t}]
\end{equation}

\vspace*{-.2cm}

\begin{figure}[t]
\vspace*{7pt}
    \centering
    \includegraphics[trim={1.5cm 0cm 3.5cm 0cm}    ,width = \linewidth]{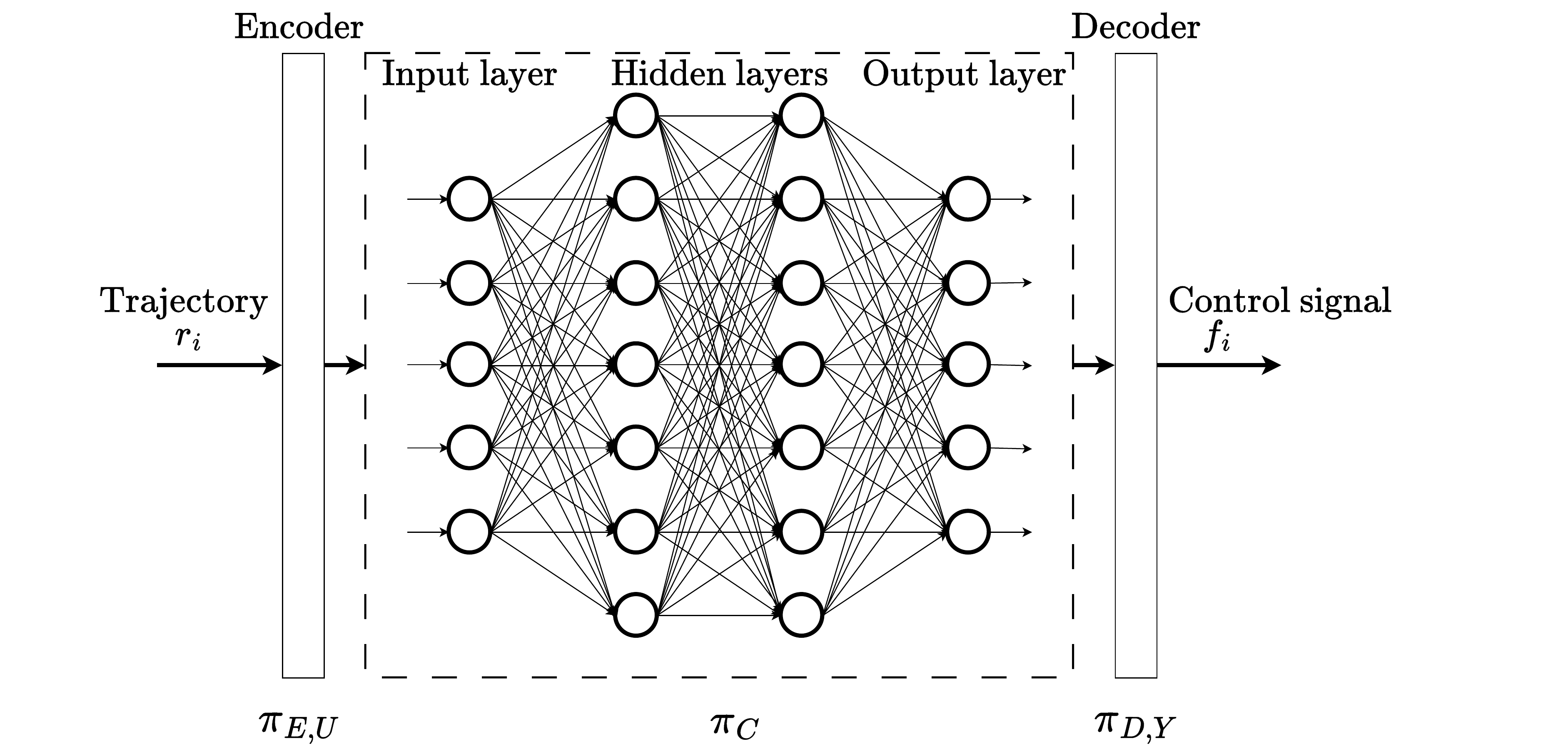} 
    \vspace*{-.6cm}
    \caption{Offline TAIL-ILC with $r_i,f_i \ \in \mathbb{R}^{n_\mathrm{d} \times 1}$.}
    \label{taililc} 
    \vspace*{-5mm}
\end{figure}

\noindent where $r_i \sim C$ and $\eta(\cdot,\cdot)$ is a performance quantification measure, and $\pi_\mathrm{s}$ are parameterized student policy candidates. The expert policy $\pi^*$ is a conventionally designed frequency domain ILC as described in Section \ref{PF}. In TAIL-ILC, the idea is to structure $ \pi_\mathrm{s}$ as :
\vspace*{-.2cm}
\begin{equation}
\label{pis}
    \pi_\mathrm{s} =  \pi_\mathrm{D,Y} \circ  \pi_\mathrm{C} \circ  \pi_\mathrm{E,U}.
\end{equation}

\vspace*{-.1cm}

\noindent
which is visualised in Figure \ref{taililc}. 
The TAIL-ILC controller is capable of generating a feedforward control signal based on a given reference trajectory. This process is carried out through a series of three sub-policies outlined in equation \eqref{pis}. The first sub-policy, $\pi_\mathrm{E,U}$, projects the reference trajectory $r_i \in \mathbb{R}^{n_\mathrm{d} \times 1}$ into a lower-dimensional space referred to as the \textit{latent space}. Next, the second sub-policy, $\pi_\mathrm{C}$, predicts a latent space representation of the feedforward signal, which is then fed into the third sub-policy, $\pi_\mathrm{D,Y}$, to project the latent space feedforward signal back into the higher-dimensional output space, resulting in $f_i \in \mathbb{R}^{n_\mathrm{d} \times 1}$. Notably, the successful application of TAIL-ILC requires that all reference trajectories share certain spatial features with each other. The prediction sub-policy, $\pi_\mathrm{C}$, is trained on a set of reference trajectories and their corresponding feedforward control signals obtained using $\pi^*$, which are projected into the latent space. The use of abstract representations enables the preservation of the most significant information of the signals while simultaneously reducing the amount of data used for making predictions, resulting in several advantages, such as increased training and prediction time efficiencies. The subsequent sub-section will delve into the development of each sub-policy in further detail.

\vspace*{-.1cm}
\subsection{Student policy $\pi_\mathrm{s}$}
The three-part student policy $\pi_\mathrm{s} : r \rightarrow f_{\pi_\mathrm{s}}$ can be decomposed into three distinct components:
\vspace{-3mm}
\begin{subequations}  
 \begin{align}
	     \pi_\mathrm{E,U} &: r \rightarrow r_l\\
	 \pi_\mathrm{C} &: r_l \rightarrow f_l\\
	 \pi_\mathrm{D,Y} &: f_l \rightarrow f_{\pi_\mathrm{s}}
  \end{align} 
     \end{subequations}
     
     \vspace*{-.1cm} \noindent
where, $r,f_{\pi_\mathrm{s}} \in \mathbb{R}^{n_\mathrm{d} \times 1}$ and $r_l,f_l \in \mathbb{R}^{n_l \times 1}$ and $n_l$ is the latent space dimensionality such that $n_l \ll n_\mathrm{d}$. As mentioned in Section \ref{3A_approach}, the training data for the sub-policy $ \pi_\mathrm{C}$, namely the pairs $\{r_{i,l},f_{i,l}\}$, are in the latent space.
\noindent This shows that the ideal outputs of $\pi_\mathrm{s}$ are of the form:
\vspace*{-.2cm}
\begin{equation}
\label{idealop}
f_{\pi_\mathrm{s}} =  \pi_\mathrm{D,Y}(f_l) = f' \approx f^*,
\end{equation}

\vspace*{-.1cm}
\noindent 
where, an approximation error may exist between $f'$ and $f^*$. Additionally, we aim at:
\vspace*{-.2cm}
\begin{equation}
 \pi_\mathrm{C}(r_l) = \widehat{f_l} \approx f_l,
\end{equation}

\vspace*{-.1cm}
\noindent 
where, in case of using a deep neural network, $\widehat{f_l}$ is the output of the network and the prediction error $e_\mathrm{pred}$ is defined as:
\vspace*{-.2cm}
\begin{equation}
	e_\mathrm{pred} = \|f_l - \widehat{f_l}\|_2,
	\end{equation}
	
	\vspace*{-.1cm}

\noindent where, $\|\cdot\|_2$ denotes the $\ell_2$ norm. Moreover, this implies that \eqref{idealop} becomes:
\vspace*{-.1cm}
\begin{equation}
	f_{\pi_\mathrm{s}} =  \pi_\mathrm{D,Y}(\widehat{f_l}) = \widehat{f'}.
\end{equation}

\vspace*{-.1cm}

\noindent In order to quantify the gap between performance of $\pi^*$ and that of $\pi_\mathrm{s}$, a distance measure is used as the performance quantification measure $\eta$ in  \eqref{mes}. This is expressed as:
\vspace*{-.1cm}
\begin{equation}
\label{newmetric}
\eta(\pi^*,\pi_\mathrm{s}) = \frac{1}{n_\mathrm{t}}\sum_{i = 1}^{n_\mathrm{t}}\|f_i - \widehat{f'_i}\|_2.
\end{equation}

\vspace*{-.1cm}

\noindent Assuming that $\mu$ represents the set of weights and biases of the deep neural network, improving the performance of $\pi_\mathrm{s}$ can be posed as the following optimization problem:
\vspace*{-.1cm}
\begin{equation}
\label{optimizationprob}
\underset{n_l,\mu}{\arg\min} \ \eta(\pi^*,\pi_\mathrm{s})
\end{equation}

\vspace*{-.1cm}

\noindent 

The proposed approach involves propagating the parameter $\eta$ through the three sub-policies, with the aim of iteratively optimizing both $n_l$ and $\mu$ via \eqref{optimizationprob}. However, given the significant computational burden associated with this approach, there is a need for a more straightforward alternative or a reformulation of the problem. With this goal in mind, we introduce the concepts of the \emph{Expert space} and \emph{Student space} to provide alternative perspectives for addressing the optimization problem at hand.

\begin{definition}
The expert space is defined as the space of all real policies denoted by superscript$\ ^\mathrm{e} $ having the form
\vspace*{-.2cm}
$$\pi^\mathrm{e}:\mathbb{R}^{n_\mathrm{x} \times 1} \rightarrow \mathbb{R}^{n_\mathrm{d} \times 1}\quad  \forall n_\mathrm{x} \in \mathbb{N}$$
\end{definition}

\noindent 
\textbf{Example:}
\begin{enumerate}
	\item Expert policy in expert space:
	\vspace*{-.2cm}
\begin{equation}\pi_\mathrm{e}^\mathrm{e}:r \rightarrow f'\end{equation}

\vspace*{-.2cm}
\noindent 
where $r,f' \in \mathbb{R}^{n_\mathrm{d} \times 1}$
	\item Student policy in expert space:
	\vspace*{-.2cm}
\begin{equation}\pi_\mathrm{s}^\mathrm{e}:r \rightarrow \widehat{f'}\end{equation} 

\vspace*{-.2cm}
\noindent 
where $r,\widehat{f'} \in \mathbb{R}^{n_\mathrm{d} \times 1}$
\end{enumerate}
\begin{definition} 
The student space is defined as the space of all real policies denoted by superscript $(\ ^\mathrm{s})$ having the form
\vspace*{-.2cm}
$$\pi^\mathrm{s}:\mathbb{R}^{n_\mathrm{x} \times 1} \rightarrow \mathbb{R}^{n_l \times 1}\ \forall \ n_\mathrm{x} \in \mathbb{N}$$
\end{definition}
\textbf{Example:}
\begin{enumerate}
	\item Expert policy in student space:
	\vspace*{-.2cm}
\begin{equation}\pi_\mathrm{e}^\mathrm{s}:r_l \rightarrow f_l\end{equation}

\vspace*{-.2cm}
\noindent 
where $r_l,f_l \in \mathbb{R}^{n_l \times 1}$
	\item Student policy in student space:
	\vspace*{-.2cm}
\begin{equation}\pi_\mathrm{s}^\mathrm{s}:r_l \rightarrow \widehat{f_l}\end{equation}

\vspace*{-.2cm}
\noindent 
where $r_l,\widehat{f_l} \in \mathbb{R}^{n_l \times 1}$
\end{enumerate}

\noindent Table \ref{policies} summarizes these definitions.

Stated differently, the expert space is comprised of all the decoding policies, $\pi_\mathrm{D}$, which project signals into $n_\mathrm{d}$ dimensions, while the student space is composed of all the encoding policies, $\pi_\mathrm{E}$, which project signals into $n_l$ dimensions.
\begin{table}[t]
\vspace{4.5pt}
\centering
    \caption{Expert and student policies in expert and student spaces}
	\begin{tabular}{|c|c|c|}
		\hline
	        &\textbf{Expert space $(^\mathrm{e})$} & \textbf{Student space $(^\mathrm{s})$}\\
		\hline
		\hline
		\textbf{Expert policy $(\pi_\mathrm{e})$} & $\pi_\mathrm{e}^\mathrm{e}:r \rightarrow f'$ & $\pi_\mathrm{e}^\mathrm{s}:r_l \rightarrow f_l$ \\
		\textbf{Student policy $(\pi_\mathrm{s})$} & $\pi_\mathrm{s}^\mathrm{e}:r \rightarrow \widehat{f'}$ & $\pi_\mathrm{s}^\mathrm{s}:r_l \rightarrow \widehat{f_l}$ \\
		\hline
	\end{tabular}
\label{policies} \vspace{-3mm}
\end{table}
Based on the preceding definitions, it is worth noting that our primary objective is to determine the student policy in the expert space, $\pi_\mathrm{s}^\mathrm{e}$. In light of these definitions, the distance metric specified in \eqref{newmetric} can be reformulated as:
\vspace*{-.2cm}
\begin{equation}
\label{modifiedupperbound}
\eta(\pi^*,\pi_\mathrm{s}^\mathrm{e}) = \eta(\pi^*,\pi_\mathrm{e}^\mathrm{e}) + \eta(\pi_\mathrm{e}^\mathrm{e},\pi_\mathrm{s}^\mathrm{e})
\end{equation}
\begin{equation*}
\label{modifiedmetric1}
\implies	\eta(\pi^{*},\pi_\mathrm{s}^\mathrm{e}) = \frac{1}{n_\mathrm{t}}\sum_{i = 1}^{n_\mathrm{t}}\|f_i - f'_i\|_2 + \frac{1}{n_\mathrm{t}}\sum_{i = 1}^{n_\mathrm{t}}\|f'_i - \widehat{f'_i}\|_2
\end{equation*}

\vspace*{-.1cm}

\noindent where, $\eta(\pi^*,\pi_\mathrm{e}^\mathrm{e})$ corresponds to the optimization of $n_l$ and $\eta(\pi_\mathrm{e}^\mathrm{e},\pi_\mathrm{s}^\mathrm{e})$ corresponds to the optimization of $\mu$. This separation of the distance measure \eqref{newmetric} allows the optimization problem in \eqref{optimizationprob} to be segmented as:
\vspace*{-.1cm}
\begin{equation*}
	\underset{n_l,\mu}{\arg\min}\ \eta(\pi^*,\pi_\mathrm{s}^\mathrm{e}) = \underset{n_l}{\arg\min}\  \eta(\pi^*,\pi_\mathrm{e}^\mathrm{e}) + \underset{\mu}{\arg\min}\ \eta(\pi_\mathrm{e}^\mathrm{e},\pi_\mathrm{s}^\mathrm{e})
\end{equation*}

\vspace*{-.1cm}

\noindent This segregation allows us to optimize $n_l$ independently of $\mu$, thus simplifying the optimization problem defined by \eqref{optimizationprob}.
\subsection{Choice of encoding and decoding sub-policies}

The encoding and decoding sub-policies in this work employ DPCA, a well-established linear dimensionality reduction technique, due to its computational simplicity. Other commonly-used linear and non-linear dimensionality reduction methods are also available and have been reviewed in \cite{osti_15002155}. 
DPCA involves the identification of a linear subspace with $n_l$ dimensions in an $n_\mathrm{d}$ dimensional space, where $n_l$ is significantly smaller than $n_\mathrm{d}$. This subspace is defined by a set of orthonormal bases that maximize the variance of the original data when projected onto this subspace. The orthonormal bases computed through this process are commonly referred to as \emph{principal components}.
\begin{definition}
A \emph{data point} in an arbitrary dataset $H \in \mathbb{R}^{n_\mathrm{x} \times n_\mathrm{t}}$ is defined as a vector $r_i \in \mathbb{R}^{n_\mathrm{x} \times 1}$ $\forall i \in [1,n_\mathrm{t}]$.
\end{definition}

The selection of the principal components for an $n_l$ dimensional latent space for the data points in $C$ involves choosing the right eigenvectors that correspond to the first $n_l$ singular values of $H$. It should be emphasized that the projection of a data point onto the latent space can be computed through the following method:
\vspace*{-.2cm}
\begin{equation}
\label{fe}
	r_l = T_\mathrm{E}  r,
\end{equation}

\vspace*{-.3cm}

\noindent where:

\vspace*{-.3cm}
\begin{equation}
\label{te}
T_\mathrm{E} = \widehat{\Sigma}^{-1} V^\top H^\top 
\end{equation}

\vspace*{-.1cm}

\noindent In this context, $r_l \in \mathbb{R}^{n_l \times 1}$, $V \in \mathbb{R}^{n_\mathrm{t} \times n_\mathrm{t}}$ denotes the matrix of right eigenvectors of $H$ and $\widehat{\Sigma} \in \mathbb{R}^{n_l \times n_\mathrm{t}}$ contains the first $n_l$ singular values of $H$ along its diagonal elements. It is worth noting that the value of $n_l$ is constrained by the number of data points in $H$. This feature of DPCA is particularly advantageous in situations where $n_{\mathrm{d}} >> n_{\mathrm{t}}$. Given the latent space representation $r_l$, a reconstructed data point $r'$ can be obtained as:
\vspace*{-.1cm}
\begin{equation}
\label{fd}
	r' = T_D  r_l, \qquad r' \in \mathbb{R}^{n_\mathrm{d} \times 1},
\end{equation} 

\vspace*{-.1cm}

\noindent 
where:
\vspace*{-.1cm}
\begin{equation}
\label{td}
T_\mathrm{D} = HV\widehat{\Sigma}^{-1}.
\end{equation}

\vspace*{-.1cm}

\begin{remark} \label{rem:2}
The computation of transformations $T_\mathrm{E}$ and $T_\mathrm{D}$ depend on $n_l$. Additionally, considering that we have access to the dataset $H$, the matrix transformations $\widehat{\Sigma}^{-1} V^TH^T$ and $HV\widehat{\Sigma}^{-1}$, the right hand side of \eqref{fe} and \eqref{fd}, become constant for a specific problem for a given choice of $n_l$. 
\end{remark}

In the light of Remark \ref{rem:2}, for a given dataset $H \in \mathbb{R}^{n_\mathrm{d} \times n_\mathrm{t}}$ the encoding ($ \pi_\mathrm{E,U}$) and decoding ($ \pi_\mathrm{D,Y}$) sub-policies for using in the student policy $\pi_\mathrm{s}$ can be defined as follows:
\vspace*{-.2cm}
\begin{subequations}
\begin{align}
\label{pie}
	 \pi_\mathrm{E,U}(r) &= T_\mathrm{E}r  \\
\label{pid}
	 \pi_\mathrm{D,Y}(\widehat{f_l}) &= T_\mathrm{D} \widehat{f_l} 
  \end{align}
\end{subequations}

\vspace*{-.1cm}

\vspace*{-.1cm}

\section{Simulation study}
\label{Section_SimulationStudy}

This section presents a comparison study of the TAIL-ILC approach in comparison with classical ILC, an artificial neural network (ANN) based ILC, referred to as NN-ILC, see \cite{Bosma-msc}, and conventional rigid body feedforward, see \cite{190Steinbuch,Proimadis-phd}, which is obtained by multiplication of the acceleration profile and the inverted rigid body dynamics of the system: 
\vspace*{-.2cm}
\begin{equation}
    C_\mathrm{FF} = m\Ddot{r_i}
\end{equation}

\begin{figure}[b]
\vspace*{-.3cm}
    \centering
    \includegraphics[width=.9\linewidth]{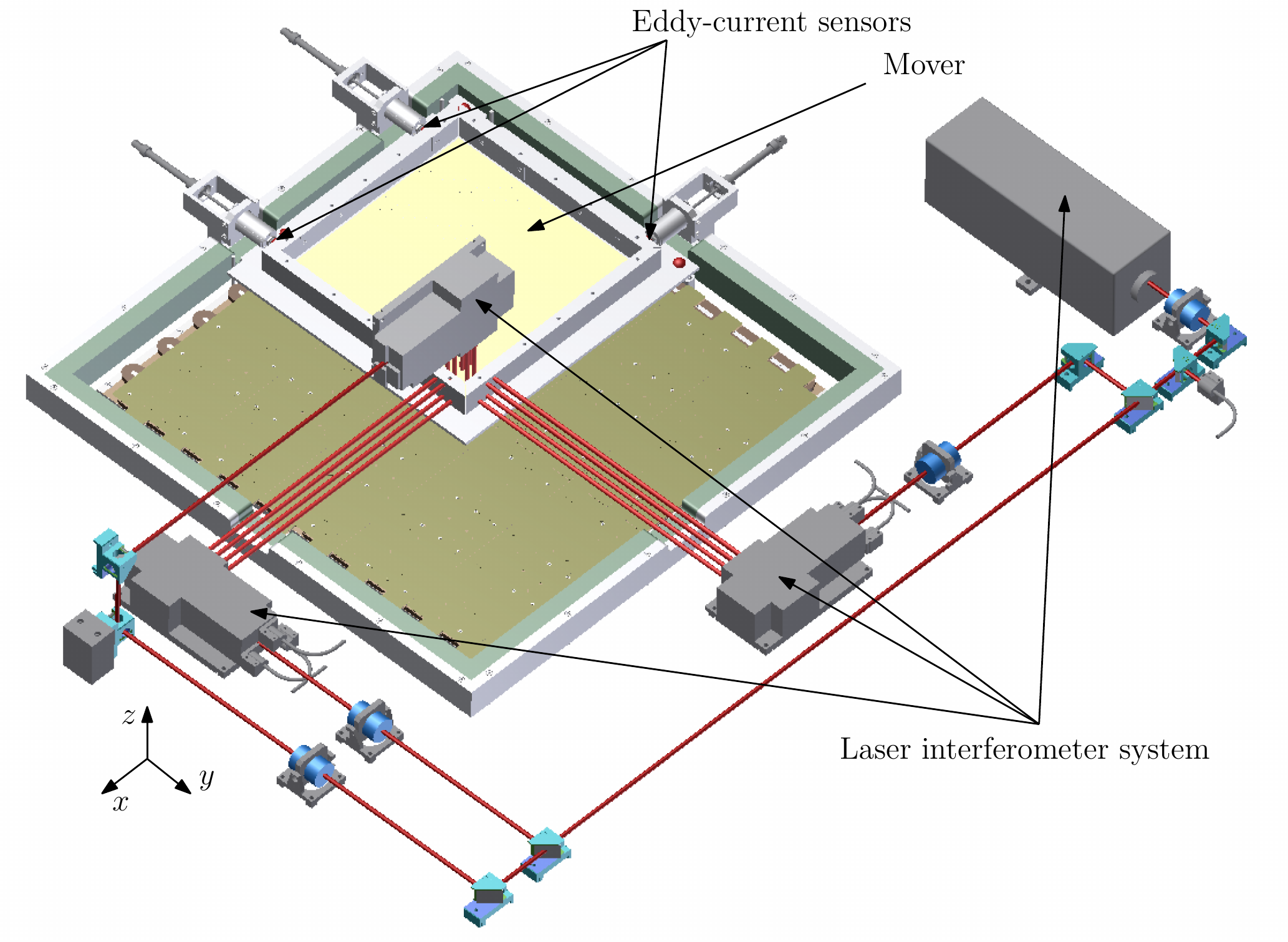}
  \vspace*{-2.5mm}
    \caption{Schematic representation of a MMPA model.}
    \label{FIGUUR}
\end{figure}

\vspace{-.1cm}
\noindent 
To facilitate simulation, a high-fidelity model of a moving-magnet planar actuator (MMPA), depicted in Figure \ref{FIGUUR}, is considered. A detailed description of a MMPA system is given in \cite{9566789}.  
Table \ref{combined-table} provide a concise overview of the network architecture and training specifics for sub-policy $\pi_\mathrm{C}$ in TAIL-ILC and policy $\pi_\mathrm{NN}$ in NN-ILC, respectively. For the sake of comparability, the training parameters are kept consistent between the two networks. The networks are designed and trained using the Deep Learning toolbox in MATLAB 2019b, employing the default random parameter initialization.

\begin{figure}[b]
\vspace*{-.6cm}
    \centering
    \includegraphics[width=\linewidth]{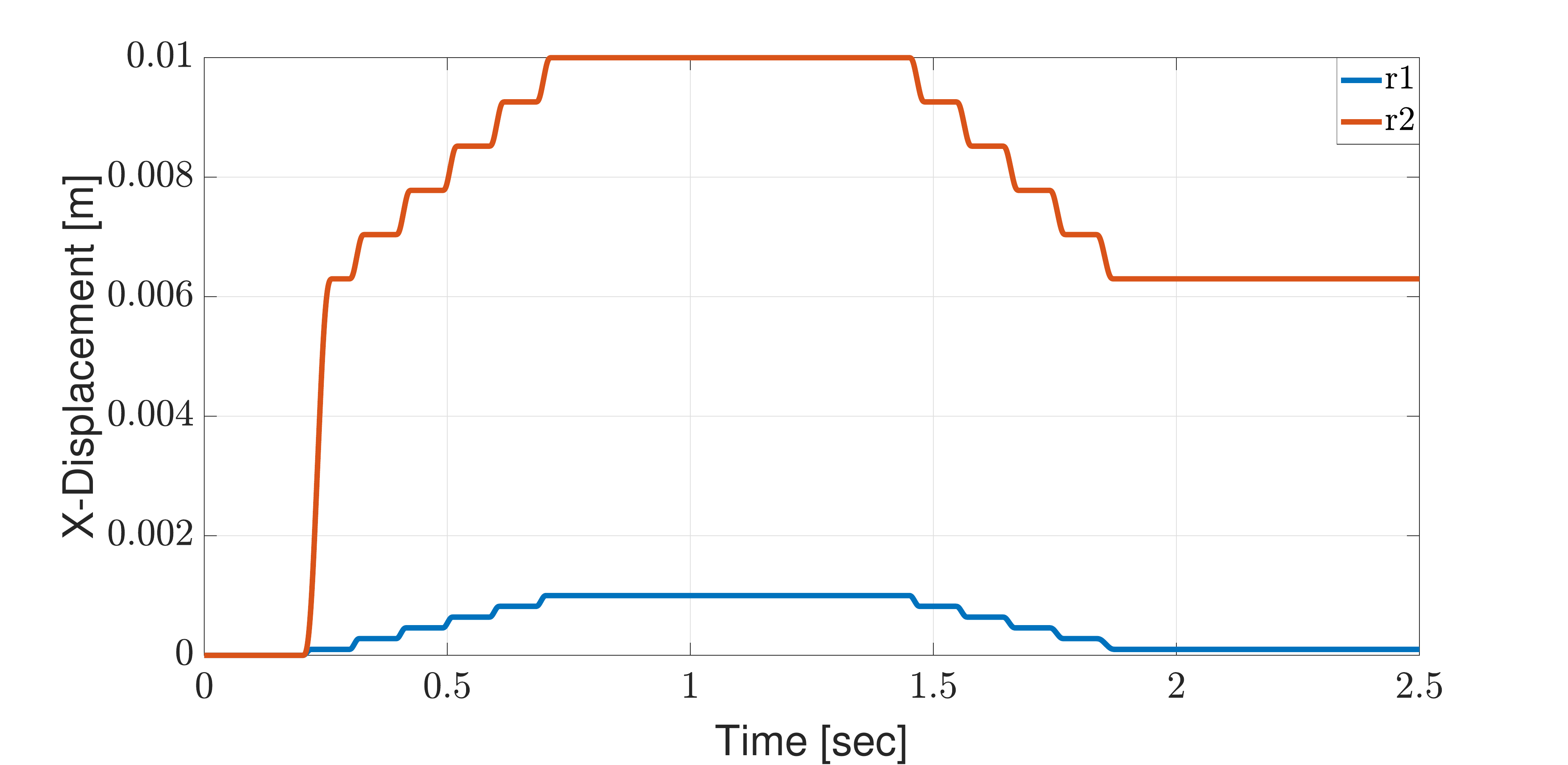} 
    \vspace*{-.6cm}
    \caption{x-direction of the references $r_1$ and $r_2$.}
    \label{reftraj1dof}
\end{figure}

The training set consists of 618 trajectories, while the test set includes 42 trajectories, each of which is 2.5 seconds long with a total of 20833 time samples. Each trajectory corresponds to a fourth-order motion profile, designed based on the approach presented by \cite{208Lambrechts}, and is parameterized with five parameters in the spatial domain. Individual trajectories are then generated by sweeping over a grid of values for each of these parameters. The objective of this study is to evaluate and compare the performance of the previously mentioned feedforward approaches against the expert ILC policy $\pi^*$, which is the traditional ILC optimized for multiple trajectories of the same class. The primary aim of ILC in this context is to mitigate any unaccounted-for residual dynamics in the system and enhance classical model-based feedforward. Consequently, we also compare the combined performance of student policies with classical feedforward controllers. We demonstrate the tracking ability of TAIL-ILC and NN-ILC on two reference trajectories, namely $r_1$ and $r_2$, which belong to the same class and are shown in Figure \ref{reftraj1dof}. $r_1$ is a randomly chosen trajectory from the training set, while $r_2$ is a previously unseen trajectory.

\begin{table}[t]
\vspace*{4.5pt}
\centering
\caption{Architecture and training details of the NNs}
\vspace*{-.2cm}
\resizebox{\linewidth}{!}{
\begin{tabular}{|c|c|c|}
\hline
\textbf{Parameter} & \textbf{TAIL-ILC} & \textbf{NN-ILC}\\
\hline
No. of neurons in the input layer & $618$ & $4$\\
\hline
No. of hidden layers & $3$ & $3$\\
\hline
No. of neurons in hidden layers & $800$ & $6$\\
\hline
Activation & Relu & Relu\\
\hline
No. of neurons in the output layer & $618$ & $1$\\
\hline
Learning rate & $10^{-3}$ & $10^{-3}$\\
\hline
Epochs & $5000$ & $5000$\\
\hline
Optimizer & adam & adam\\
\hline
Minibatch size & $128$ & $128$\\
\hline
Train set & $618$ trajectories & $618$ trajectories\\
\hline
Test set & $42$ trajectories & $42$ trajectories\\
\hline
\end{tabular}}
\vspace*{-.4cm}
\label{combined-table}
\end{table}

\vspace{-.1cm}
\subsection{Time domain performance of TAIL-ILC and NN-ILC}
\begin{figure}[t]
\vspace*{4.5pt}
\centering
\begin{subfigure}{.5\linewidth}
  \centering
  \includegraphics[trim={1cm 0.5cm .2cm 0cm}
    ,width=\linewidth]{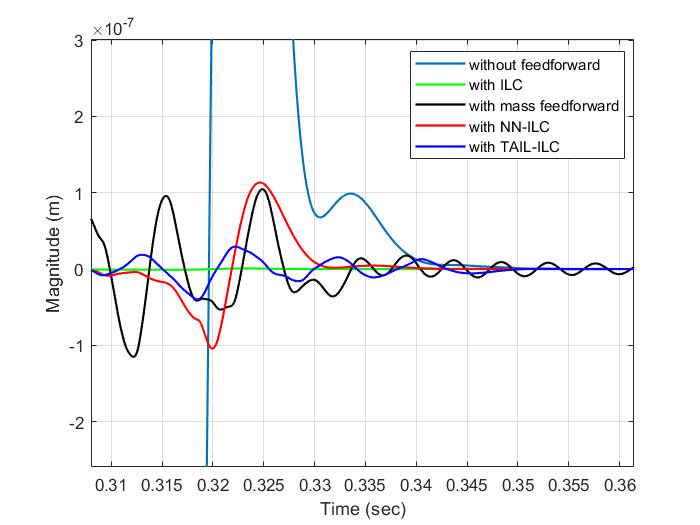}
\end{subfigure}%
\begin{subfigure}{.5\linewidth}
  \centering
  \includegraphics[trim={1cm 0.5cm .2cm 0cm}
    ,width=\linewidth]{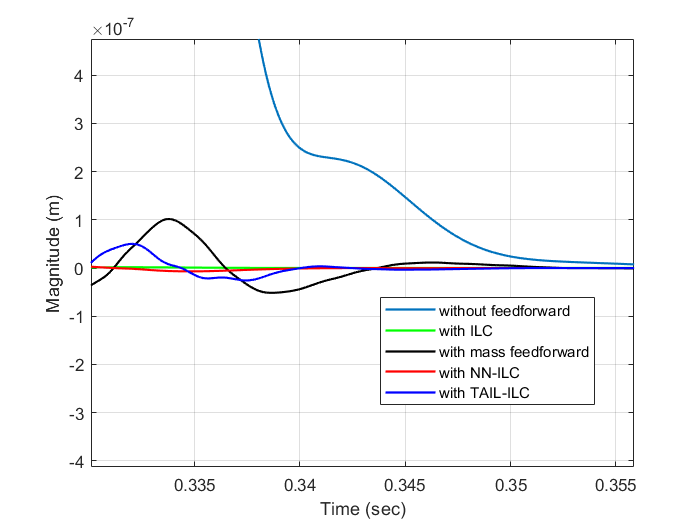}

\end{subfigure}
\caption{{Tracking error for $r_1$ (left) and $r_2$ (right) during constant velocity.}}
\label{trkerrseen}
\vspace*{-.5cm}
\end{figure}

A silicon wafer scanning application is considered where the scanning takes place during the constant velocity interval of the motion profile, see \cite{Butler}. In this context, Figure \ref{trkerrseen} illustrates the position tracking error in $x$-direction during the constant velocity interval of the reference trajectories $r_1$ and $r_2$ respectively. In addition to the performance of mass feedforward, TAIL-ILC and NN-ILC, the figure also indicates the performance of the expert ILC policy. This is to facilitate the comparison of the performance of the two deep learning based ILC variants with the baseline. As demonstrated in the left Figure, i.e. the performance of the feedforward controllers on $r_1$, the expert ILC policy exhibits the highest overall performance. Nonetheless, it is noteworthy that the TAIL-ILC policy outperforms in terms of the peak tracking error achieved compared to the alternative feedforward approaches, whereas the NN-ILC policy demonstrates a superior performance in terms of the convergence time of the error. Nonetheless, when analyzing the right Figure, i.e. the performance of the feedforward approaches for a previously unseen trajectory $r_2$, the expert ILC policy needs to re-learn the relevant feedforward signal. Conversely, the TAIL-ILC and NN-ILC policies are capable of achieving similar performance to the re-learned expert ILC policy without any further training. Additionally, when combined with a classical mass feedforward controller, both the TAIL-ILC and NN-ILC policies are observed to yield superior performance in terms of peak error and settling time compared to the classical mass feedforward controller alone.



\vspace{-.1cm}
\subsection{TAIL-ILC vs NN-ILC}
Table \ref{comparison results} provides a comparison of the training and prediction properties of the TAIL-ILC and NN-ILC student policies. Here, we compare the following parameters:
\begin{enumerate}
    \item $T_\text{train}$: Time to train the neural network
    \item $T_\text{predict}$: Time to make predictions for 10 randomly selected test set trajectories.
    \item $e_\text{train}$: Control signal prediction error averaged over 10 randomly selected train set trajectories.
	\item $e_\text{test}$: Control signal prediction error averaged over 10 randomly selected test set trajectories.
	\item $e_\text{peak \ tracking}$: Peak tracking error achieved with the predicted control signals averaged over 10 randomly selected train set trajectories.
\end{enumerate}
\begin{table}[b] 
\vspace*{-.3cm}
		\centering
		\caption{Performance comparison for the $1^{st}$ degree of freedom}
		\begin{tabular}{|c|c|c|}
			\hline
			\textbf{Criterion} & \textbf{NN-ILC} & \textbf{TAIL-ILC}\\
			\hline
	        $T_\text{train}$ & $2.5$ hr & $20$ min\\
	        \hline
			$T_\text{predict}(\text{per sample})$ & $0.005$ sec & $0.064$ sec\\
			\hline
			$T_\text{predict}(\text{full signal})$ & $86$ sec & $0.064$ sec\\
			\hline
			$e_\text{train}$ & $0.0055$ N & $0.0011$ N\\
			\hline
			$e_\text{test}$ & $0.0013$ N & $0.0064$ N\\
			\hline
			$e_\text{peak\ tracking}$ & $1.3 \cdot 10^{-7}$ m & $8.3 \cdot 10^{-8}$ m\\
			\hline
		\end{tabular}
		\label{comparison results} 
\end{table}
\noindent Here, the average control signal prediction errors of the train and the test set trajectories are calculated as the values of the performance measure $\eta$ in \eqref{modifiedupperbound}. As can be seen, though the original signals and trajectories are extremely high dimensional, the projection of these signals into the latent space using the proposed TAIL-ILC approach has resulted in significant improvement in training and prediction time compared to that of the NN-ILC approach.  


Moreover, as observed in Figure \ref{trkerrseen}, the average signal prediction error has decreased for TAIL-ILC in case of previously seen trajectories whereas the NN-ILC has improved performance for previously unseen trajectories.

\vspace{-.2cm}
\section{TAIL-ILC vs NN-ILC PERSPECTIVES}
\label{TAILPERSPE}

\vspace*{-.1cm}
In the previous Section, we have seen a comparison of the performance of TAIL-ILC and NN-ILC controllers for a specific use case. However, it is more natural to view these controllers as individual instances of two fundamentally different perspectives of the problem. Hence, it is important to reflect upon the perspectives that these controllers convey and the consequences for various aspects of the resulting controllers. This is expected to provide us with a more generalised reasoning to some of the differences observed in performances of these two controllers.

\vspace{-.1cm}
\subsection{Time duration of trajectories}

\vspace*{-.1cm}

The NN-ILC and TAIL-ILC are two approaches of ILC that differ in their treatment of reference trajectories and feedforward signals. NN-ILC is capable of handling trajectories of different lengths, as it deals with them sample-wise. In contrast, TAIL-ILC processes trajectories and signals in their entirety, making it challenging to manage trajectories of varying durations due to the fixed input-output dimensionality of neural network learning models. Additionally, NN-ILC is better equipped to handle instantaneous changes in reference trajectories compared to TAIL-ILC. A possible solution to reconcile these perspectives is to use a different class of learning models, such as a recurrent neural network.
\subsection{Training and prediction time efficiencies}
\vspace*{-.1cm}

In NN-ILC, the training dataset used for $\pi_\mathrm{NN}$ encompasses all the samples from all the trajectories in the training set, along with their associated feedforward signals. Conversely, TAIL-ILC employs a training dataset for $\pi_\mathrm{C}$ that solely includes the parameters of the trajectories and feedforward signals within the latent space, resulting in a significantly smaller dataset in comparison to the total number of samples. This characteristic leads to TAIL-ILC presenting shorter training and prediction times when compared to NN-ILC, as demonstrated by the results presented in Table \ref{comparison results}.
\subsection{Generalizability to previously unseen trajectories}

Figure \ref{trkerrseen} demonstrates that NN-ILC outperforms TAIL-ILC in terms of generalizing performance to previously unobserved trajectories. The improved performance can be attributed to NN-ILC's treatment of reference trajectories as points in an $n$-dimensional space corresponding to $n^\mathrm{th}$ order motion profiles, which allows it to learn a mapping to the corresponding feedforward signal time samples. As a result, the trained network can more accurately extrapolate performance to previously unobserved points in the space of possible reference trajectories. In contrast, TAIL-ILC relies primarily on analogies between individual tasks on a higher level, which may result in suboptimal performance when confronted with previously unobserved trajectories at the sample level.
\vspace{-.1cm}
\section{CONCLUSION}
\label{Section_Conclusions}
\vspace*{-.1cm}

In this work, we have primarily explored two different perspectives within the context of deep learning of the task-flexibility constraint of conventional ILC. While each of the considered approaches has its own advantages and disadvantages, it has been observed that the use of deep learning techniques in general could be a useful direction for future research in designing task-flexible ILC variants.

\vspace*{-.1cm}

\bibliographystyle{ieeetr}       
\bibliography{MyBib}

\end{document}